\documentstyle[12pt]{article}
\@addtoreset{equation}{section}
\def\theequation{\arabic{section}.\arabic{equation}}

\def\section{\@startsection{section}{1}{\z@}{3.5ex plus 1ex minus
   .2ex}{2.3ex plus .2ex}{\large\bf}}
\def\thesection{\Roman{section}.}

\def\appendix{\setcounter{section}{0}
        \def\thesection{Appendix\ \Alph{section}.}
        \def\theequation{\Alph{section}.\arabic{equation}}}

\newcommand{\beq}{\begin{eqnarray}}
\newcommand{\eeq}{\end{eqnarray}}
\newcommand{\eq}{eqnarray}

\newcommand{\al}{{\alpha}}
\newcommand{\be}{{\beta}}
\newcommand{\ci}{\cite}
\newcommand{\ga}{{\gamma}}
\newcommand{\Ga}{{\Gamma}}
\newcommand{\ep}{{\epsilon}}

\newcommand{\de}{{\delta}}

\newcommand{\la}{{\lambda}}
\newcommand{\La}{{\Lambda}}
\newcommand{\m}{{\mu}}
\newcommand{\n}{{\nu}}
\newcommand{\si}{{\sigma}}

\newcommand{\pa}{{\partial}}
\newcommand{\no}{{\nonumber}}
\newcommand{\f}{\frac}
\newcommand{\ra}{\rightarrow}
\newcommand{\lra}{\leftrightarrow}

\newcommand{\hb}{\hat{\be}}

\newcommand{\diff}{diffeomorphism }

\begin{document}
\topmargin 0pt \oddsidemargin -3.5mm \headheight 0pt \topskip 0mm
\addtolength{\baselineskip}{0.20\baselineskip}
\begin{flushright}
   arXiv:0705.4381[hep-th]
\end{flushright}
\vspace{0.1cm}
\begin{center}
  {\large \bf Holography in Three-dimensional Kerr-de Sitter
  Space with a Gravitational Chern-Simons Term }
\end{center}
\vspace{0.1cm}
\begin{center}
 Mu-In Park\footnote{Electronic address:
muinpark@gmail.com}
\\{
Research Institute of Physics and
Chemistry}, \\{Chonbuk National University, Chonju 561-756, Korea} 
\\
\end{center}
\vspace{0.1cm}
\begin{center}
  {\bf ABSTRACT}
\end{center}

The holographic description of the three-dimensional Kerr-de Sitter
space with a gravitational Chern-Simons term is studied, in the
context of dS/CFT correspondence. The space has only one
(cosmological) event horizon and its mass and angular momentum are
identified from the holographic energy-momentum tensor at the
asymptotic infinity. The thermodynamic entropy of the cosmological
horizon is computed directly from the first law of thermodynamics,
with the conventional Hawking temperature, and it is found that the
usual Gibbons-Hawking entropy is modified.
It is remarked that, due to the gravitational Chern-Simons term, (a)
the results go beyond the analytic continuation from AdS, (b) the
maximum-mass/{\bf N}-bound conjecture may be violated, and (c) the
three-dimensional cosmology is chiral.
A statistical mechanical computation of the entropy, from a
Cardy-like formula for a dual CFT at the asymptotic boundary, is
discussed. Some technical difference in the Chern-Simons
energy-momentum tensor, from literatures is remarked also.

 \vspace{0.1cm}
\begin{flushleft}
\\14 May 2008 \\
\end{flushleft}
\newpage

\begin{section}
{Introduction}
\end{section}

In the last several years, the holographic description of
asymptotically anti-de Sitter (AdS) spaces has been extensively
studied in the context of AdS/CFT correspondence \cite{Ahar:99}. On
the other hand, recently, some higher derivative corrections to the
description have been studied also \cite{Noji:99,Noji:01,Cvet:01},
and it is found that, when there are black holes in the bulk
space-times, there are good agreements even with the higher
derivative corrections to the black hole entropies
\ci{Moha:05,Said:99,Krau:05b,Park:0609}.

More recently, the corrections due to a gravitational Chern-Simons
term \ci{Dese:82} have been studied also, and it is found that, when
considering the three-dimensional AdS space with a negative
cosmological constant $\Lambda=-1/l^2$, having a black hole, known
as the BTZ black hole \ci{Bana:92}, the black hole's mass and
angular momentum are modified as
\begin{eqnarray}
M&=&m+\hb j/l, ~J=j+ \hb lm, \label{MJ:general}
 \label{S:old}
\end{eqnarray}
which shows some $mixings$ between the original BTZ black hole's
mass $m$ and angular momentum $j$, with the coupling constant $\hb$
of the gravitational Chern-Simons term\footnote{Similar phenomena
have been known, for sometime, in several other context where the
masses and angular momenta are {\it completely} interchanged
\ci{Carl:91,Carl:94,Bana:98a,Bana:98c}.}. This has been computed in
some standard canonical methods
\ci{Garc:03,Mous:03,Blag:04,Dese:05,Olme:05} and in the holographic
methods \ci{Krau:05a,Solo:05a}. It is found also that the black hole
entropy which satisfies the first law of thermodynamics with the
mass and angular momentum of (\ref{MJ:general}) has a term
proportional to the inner-horizon's area, as well as the usual
outer-horizon's \ci{Solo:05a,Saho:06,Tach:06}. Moreover, the entropy
agrees with the statistical entropy, computed from the Cardy's
formula for a CFT at the asymptotic infinity
\ci{Krau:05a,Solo:05a,Saho:06,Park:0602,Park:0608}.

In this paper, I consider the three-dimensional Kerr-de Sitter space
(KdS$_3$) with a gravitational Chern-Simons term, in the context of
dS/CFT correspondence, as a de Sitter counter part of the above
mentioned analysis
\ci{Park:9806,Park:9811,Stro:0106,Mazu:01,Noji:01b,Klem:01,Spra:01,
Bala:01,Myun:01,Cai:01,Ghez:01}. As far as the higher ``curvature''
corrections, there are several works already in the literatures in
the dS/CFT \ci{Noji:99,Cvet:01}. But, the analysis with the
Chern-Simons term, in particular, is
interesting from the following reasons. First, the usual analytic
continuation \ci{Bala:01,Sken:02} is questionable from the different
behaviors of the Einstein action and the gravitational Chern-Simons
term under the continuation \ci{Witt:88}: One gets a {\it real}
(-valued) action for the former but
an {\it imaginary} for the latter so that
one can {\it not}
get
a real
total action but a {\it complex} action. This implies that the AD
mass,
angular momentum \ci{Dese:05,Olme:05}, and
the entropy be {\it complex} by the formal analytic continuation.
Second, it is known recently that the entropic ${\bf N}$-bound
\ci{Bous:99,Bous:00}, stating that {\it the upper bound of the
entropy in asymptotically de Sitter space is given by the entropy of
pure de Sitter space}, and the maximal mass conjecture
\ci{Bala:01,Cai:01,Ghez:01,Ghez:05}, stating that {\it any
asymptotically de Sitter space can not have a mass larger than the
pure de Sitter case without inducing a cosmological singularity},
may be violated with NUT-charge, where the entropy is no longer
proportional to the area \ci{Clar:03,Clar:04,Mann:04}. However, in
our case with the gravitational Chern-Simons term also, one would
have an entropy which is not proportional to the (outer-horizon)
area \ci{Solo:05a,Saho:06,Tach:06}, in the real action approach
\ci{Olme:05} without recourse to the analytic continuation. So, it
would be interesting to study whether the two conjectures are
violated similarly in our case also. Third, the Chern-Simons term
introduces chirality into cosmology in four dimensions which could
produce some interesting effects in our real world
\ci{Lue:98,Jack:03,Lyth:05}. It would be interesting to study the
corresponding effects in the three-dimensional cosmology model.
Finally, it is known \ci{Park:9806,Bala:01} that the Gibbons-Hakwing
entropy for the cosmological horizon agrees with the statistical
entropy, computed from the Cardy-like formula at the infinite
boundary, even though the spacetime gives rise to a {\it
non-unitary} CFT, due to {\it complex} eigenvalues for the Virasoro
generators $L_0, \bar{L}_0$, and so the formula does not generally
apply. It would be interesting to explore how much this Cardy
formula approach can be generalized further with the quite
non-trivial modification of the Einstein gravity theory.

The organization of this paper is as follows. In Sec. II, I review
the holographic energy-momentum tensor, at the asymptotic infinity,
for Einstein gravity in dS space. In Sec. III, I describe the
gravitational Chern-Simons term, in the real action approach, and
compute its contribution to the holographic energy-momentum tensor.
In Sec. IV, I consider the Fefferman-Graham expansion and identify
the boundary energy-momentum tensor. It is noted also that the
Chern-Simons contributions go beyond the analytic continuation. In
Sec. V, I consider KdS$_3$ space, compute its holographic
energy-momentum tensor, and identify the conserved mass and angular
momentum of the space. I consider the entropy of the cosmological
horizon in the KdS$_3$ space from the first law of thermodynamics,
with the conventional Hawking temperature. It is remarked also that,
due to the Chern-Simons term, the maximum-mass/{\bf N}-bound
conjecture may be violated and the three-dimensional cosmology is
chiral. In Sec. VI, I consider a statistical computation of the
entropy from a Cardy-like formula for a dual CFT at the asymptotic
boundary. In Appendix {\bf A}, I present some details about the
Gauss-Godazzi equations and their Fefferman-Graham expansions, by
comparing the cases of the dS and AdS spaces. In Appendix {\bf B}, I
present some details about computing the gravitational Chern-Simons
contributions to the energy-momentum, to clarify some differences in
the details with the literatures. In Appendix {\bf C}, I present the
comparative computations of the conserved charges for dS and AdS
cases to clarify some differences in the details from the
literatures. It is remarked also that, due to the gravitational
Chern-Simons term, the results go beyond the analytic continuation
from AdS.

I shall omit the speed of light $c$ and the Boltzman's constant
$k_B$ in this paper for convenience, by adopting the units of
$c\equiv 1,~k_B \equiv 1$. But, I shall keep the Newton's constant
$G$ and the Planck's constant $\hbar$ in order to clearly
distinguish the quantum (gravity) effects with the classical ones.\\

\begin{section}
{Holographic energy-momentum tensor in dS space: Einstein gravity
case}
\end{section}

Brown and York \ci{Brow:93} have given a general description for
defining an energy-momentum tensor, associated to a boundary
$\partial {\cal M}$ of a spacetime ${\cal M}$. In order to study the
asymptotic infinite boundary in three-dimensional, asymptotically dS
spacetime with a positive cosmological constant $\La=1/l^2$, let me
slice the spacetime\footnote{Greek letters $(\m, \n, \cdots)$ denote
the three-dimensional indices, whereas Roman letters $(i, j,
\cdots)$ denote the two-dimensional boundary indices.}
\begin{\eq}
\label{slice}
 ds^2=g_{\m \n} dx^{\m} dx^{\n}=-d \eta^2
+\ga_{ij}(\eta, x^i) dx^i dx^j
\end{\eq}
with two-dimensional hypersurfaces labeled by $\eta$. Then, the
Einstein-Hilbert action, accompanied by the extrinsic curvature term
on the boundary $\partial {\cal M}$ \ci{York:86}, is
\begin{eqnarray}
\label{EH} I_{g}&=&\frac{1}{16 \pi G} \int_{\cal M} d^3 x \sqrt{|g|}
\left( {}^{(3)}R -2 \La \right)-\frac{1}{8 \pi G} \int_{\pa {\cal
M}}
d^2 x \sqrt{|\ga|} Tr K \no \\
 &=&\frac{1}{16 \pi G} \int_{\cal M}
d^2 x d\eta  ~ \sqrt{|\ga|} \left( R+(Tr K)^2 +Tr (K^2)-2 \La
\right).
\end{eqnarray}
Here, I introduced the extrinsic curvature of a fixed-$\eta$
surface, $K_{ij}=\pa_{\eta} \ga_{ij}/2$ and used the decomposition
of the three-dimensional Ricci scalar ${}^{(3)}R$, in terms of the
two-dimensional Ricci scalar of the metric $\ga_{ij}$,
\begin{\eq}
{}^{(3)}R=R+ (Tr K)^2 +Tr (K^2) +2 \pa_{\eta} Tr K,
\end{\eq}
where $Tr K=\ga^{ij} K_{ij},~Tr(K^2)=K^{ij} K_{ij}$\footnote{I
follow the conventions of Wald \ci{Wald:84}, i.e., ${}^{(3)}{R_{\m
\n \be}}^{\al}=\pa_{\nu} \Ga_{\be \mu}^{\al}+\Ga_{\ga \nu}^{\al}
\Ga_{\be \mu}^{\ga}-(\n \lra \m), \Ga_{\m \n}^{\la}=({1}/{2}) g^{\la
\si} (\pa_{\m}g_{\n \si} +\pa_{\n} g_{\si \mu}-\pa_{\si} g_{\m \n}
), {}^{(3)} R_{\m \n} = {}^{(3)} {R_{\m \la \n}}^{\la}, {}^{(3)} R =
{}^{(3)} {R_{\m}}^{\m}$. These conventions agree with those in Refs.
\ci{Stro:0106,Klem:01,Ghez:01,Noji:99,Brow:93,York:86,Krau:99}, but
differ from those in Refs. \ci{Solo:05a,Bala:01,Bala:99}, in the
sign of the Riemann tensor.}.

To compute the energy-momentum tensor let me consider the variation
of the action with respect to metric. In general, the variation
produces bulk terms proportional to the equations of motion plus
some boundary terms:
\begin{\eq}
\de I_g =-\frac{1}{16 \pi G} \int_{\cal M} d^3 x \sqrt{|g|} \left(
{}^{(3)}R -\frac{1}{2}   g^{\m \n}~ {}^{(3)} R +\f{1}{l^2} g^{\m \n}
\right) \de g_{\m \n} +(\mbox{boundary terms}).
\end{\eq}
But, if one considers the solutions to the {\it bulk} equations of
motion,
\begin{\eq}
\label{eom:bulk} {}^{(3)}R^{\m \n} -\frac{1}{2} g^{\m \n}~ {}^{(3)}
R +\f{1}{l^2} g^{\m \n} =0,
\end{\eq}
only the boundary terms remain as follows \ci{Brow:93}:
\begin{\eq}
\label{dEH:on-shell} \de I_{g(on-shell)} =\frac{1}{2} \int_{\pa{\cal
M}} d^2 x \sqrt{|\ga|} T^{ij} \de \ga_{ij}.
\end{\eq}
Here, the Brown-York's boundary energy-momentum tensor is given by
\begin{\eq}
T^{ij} =\f{1}{8 \pi G} (K^{ij} -Tr K  \ga^{ij}   ).
\end{\eq}
Since I want to think of the boundary at $\eta=\infty$ with a finite
energy-momentum tensor \ci{Bala:99,Krau:99,Stro:0106,Mazu:01}, I
need to consider additional counter terms also which can be fixed by
the locality and general covariance,
\begin{\eq}
\label{I:ct}
 I_{ct}=\frac{1}{8 \pi G l} \int_{\pa{\cal M}} d^2 x
\sqrt{|\ga|}.
\end{\eq}
Then, the regulated energy-momentum tensor becomes
\begin{\eq}
\label{Tij:reg}
 T^{ij}_{reg}=T^{ij}+\f{1}{8 \pi G l} \ga^{ij}.
\end{\eq}
\\
\begin{section}
{Gravitational Chern-Simons term and its contribution to the
energy-momentum tensor}
\end{section}
The gravitational Chern-Simons term, in the ${form}$ notation
\ci{Gins:85,Krau:05a}, is given by
\begin{\eq}
 I_{GCS} =\be_{KL} \int_{\cal M} Tr  \left( \Ga \wedge d \Ga
+\f{2}{3} \Ga \wedge \Ga \wedge \Ga \right),
\end{\eq}
where I have defined the connection 1-form as
$\Ga_{\be}^{\al}=\Ga_{\be \mu}^{\al} dx^{\mu}$, with the usual
Christoffel symbols $\Ga_{\be \mu}^{\al}$, and $\be_{KL}$ is the
{\it real}-valued coupling which agrees with that of Ref.
\ci{Krau:05a}. Note that $I_{GCS}$ is of the third-derivative order,
rather than the second as in the Einstein-Hilbert term in
(\ref{EH}), and so it is the first higher-derivative correction in
three-dimensional spacetimes. Moreover, it has a peculiar property
which differs from that of the higher ``curvature'' corrections: The
action is not manifestly invariant under the diffeomorphism but its
bulk equations of motion are covariant \ci{Gins:85}; this implies
that the non-invariance of \diff propagates only along the boundary
and this introduces a gravitational anomaly on the boundary
\ci{Krau:05a}.

Now, in order to compute the contribution to the energy-momentum
tensor, let me consider the variation of the action $I_{GCS}$ with
respect to metric. After some computation (see the Appendix {\bf B}
for some details), one can find that
\begin{\eq}
\label{dGCS} \be_{KL}^{-1} \de I_{GCS} &=&2 \int_{\cal M} Tr  \left(
\de \Ga \wedge R\right)-\int_{\pa {\cal M}} Tr  \left( \Ga \wedge
\de \Ga \right) \no
\\
&=&-\int_{\cal M} d^3 x \nabla_{\be} {}^{(3)} {R_{\ga \rho}}^{\m
\be}
\ep^{\n \rho \ga} \de g_{\m \n}  \no \\
&&+ \int_{\pa{\cal M}} d^2 x \left[ {}^{(3)} {R_{\nu \mu}}^{i \eta}
\ep^{j \m \n} \de \ga_{ij} +(-2 {K^{l}}_i \de K_{lj}-\Ga_{li}^k \de
\Ga_{kj}^l) \ep^{ij}\right],
\end{\eq}
where the curvature 2-form is given by ${}^{(3)}{R_{\be}}^{\al} =d
{\Ga_{\be}}^{\al} +\Ga^{\al}_{\ga} \wedge \Ga^{\ga}_{\be} =({1}/{2})
{}^{(3)}{R_{\n \m \be}}^{\al} dx^{\m} \wedge dx^{\n}$ with the usual
Riemann tensor ${}^{(3)}{R_{\n \m \be}}^{\al}$, and $\nabla_{\be}$
denotes the covariant derivative with respect to $g_{\al \be}$. And
also, I have used $dx^{\ga} \wedge dx^{\m} \wedge dx^{\n} =\ep^{\ga
\m \n} d^3 x,~ dx^i \wedge dx^j =\ep^{ij} d^2 x$ with
$\ep^{012}\equiv 1, \ep^{01} \equiv 1 $ \footnote{The result
(\ref{dGCS}) differs from that of Ref. \ci{Krau:05a} by the absence
of $\sqrt{ |\gamma| }$ factor, but agrees with that of Ref.
\ci{Dese:82}. From this difference, there follows the difference in
the energy-momentum tensor $t^{ij}$ of (\ref{tij}) by the factor of
$1/\sqrt{| \ga | }$, from that of literatures
\ci{Krau:05a,Solo:05a}.}. Note that the bulk term is a manifestly
covariant form and this gives a covariant contribution to the
equations of motion (\ref{eom:bulk}) as follows:
\begin{\eq}
\label{eom:GCS} {}^{(3)}R^{\m \n} -\frac{1}{2} g^{\m \n} ~{}^{(3)} R
+\f{1}{l^2} g^{\m \n} =-16 \pi \be_{KL} \ep^{\n \rho \ga}
\nabla_{\be} {}^{(3)}{R_{\ga \rho}}^{\m \be}/\sqrt{|g|}.
\end{\eq}
Now, by considering the solution to the full equations of motion for
the action, in the presence of the gravitational Chern-Simons term
$I_{GCS}$ as well as the Einstein-Hilbert term $I_g$ of (\ref{EH}),
the boundary term in (\ref{dGCS}) would contribute to the boundary
energy-momentum tensor $t^{ij}$ as follows
\begin{\eq}
\label{dGCS:on-shell} \de I_{GCS(on-shell)} =\frac{1}{2}
\int_{\pa{\cal M}} d^2 x \sqrt{|\ga|} t^{ij} \de \ga_{ij}
\end{\eq}
with
\begin{\eq}
\label{tij} t^{ij}=\f{2}{\sqrt{|\ga|}} \be_{KL} {}^{(3)} {R_{\n
\m}}^{i \eta} \ep^{j  \m \n} +\cdots ,
\end{\eq}
where `$\cdots$' denotes the contributions, if there are, from the
second term in the boundary terms in (\ref{dGCS}). By just a naive
looking at the second boundary term in (\ref{dGCS}), it is not clear
whether one can define the local quantity through
(\ref{dGCS:on-shell}) since $\de K_{lj}$ and $\de \Ga_{kj}^l$
involve some derivatives of $\de \ga_{ij}$. Generally, we need to
introduce some appropriate boundary terms in order to compensate
these unwanted derivative terms, but these appropriate boundary
terms are known only for some limited cases, like as the
Einstein-Hilbert action or its modification by the Gauss-Bonnet term
\ci{Myer:87}. But, fortunately, in our case of asymptotically dS (or
AdS) space-times these are not needed for an infinite boundary at
$\eta=\infty$, as one can see explicitly in the next section.\\

\begin{section}
{Fefferman-Graham expansion and boundary energy-momentum tensor}
\end{section}

In order to study the physics  associated with the asymptotic
boundary at $\eta=\infty$, it is convenient to consider the large
$\eta$ expansion, known as the Fefferman-Graham expansion
\ci{Feff:85}, of the boundary metric $\ga_{ij}(\eta, x^i)$ as
follows \footnote{For higher-dimensional (A)dS space, $(A)dS_{2n+1}$
with $n \ge 2$, there is also the $log (\ep)$ term generally
\ci{Henn:98,Sken:02}. But, this is not needed in the
three-dimensional case.}:
\begin{\eq}
\label{FG}
 \ga_{ij}(\eta, x^i)=\ep^{-1}
\ga^{(0)}_{ij}+\ga^{(2)}_{ij}+\ep \ga^{(4)}_{ij} +O (\ep^2)
\end{\eq}
with $\ep=e^{-2 \eta/l}$. Then, it follows that
\begin{\eq}
\label{K:FG}
 \ga^{ij}&=&\ep \ga^{(0)ij}-\ep^2 \ga^{(2) ij}+\ep^3 [
(\ga^{(2)})^2-\ga^{(4)}]^{ij} +O (\ep^4), \no \\
K_{ij}&=&\f{1}{l} \left[ \ep^{-1} \ga^{(0)}_{ij}-\ep \ga^{(4)}_{ij} \right], \no \\
{K^{i}}_{j}&=&\f{1}{l} \left[ {\ga^{(0)i}}_{j}- \ep {\ga^{(2)
i}}_{j}+\ep^2 {[
(\ga^{(2)})^2-\ga^{(4)}]^{i}}_{j}  \right] +O (\ep^3), \no \\
K^{ij}&=&\f{1}{l} \left[ \ep \ga^{(0)ij}-2 \ep^2 \ga^{(2)ij}+ \ep^3
[
3 (\ga^{(2)})^2-2 \ga^{(4)}]^{ij} \right] +O (\ep^4) , \no \\
TrK&=&\f{1}{l} Tr \left[\ga^{(0)}-\ep \ga^{(2)}+\ep^2 [
(\ga^{(2)})^2-\ga^{(4)}\right] +O (\ep^3) ,
\end{\eq}
and one can
obtain the energy-momentum tensor for the Einstein-Hilbert action as
follows:
\begin{\eq}
\label{Tij:reg:FG}
 T^{ij}_{reg} =-\f{1}{8 \pi G} \f{\ep^2}{l} \left(
\ga^{(2)ij}- \ga^{(0)ij}  Tr \ga^{(2)}  \right) +O(\ep^3).
\end{\eq}
But, this naively defined quantity vanishes as $\ep \ra 0$, i.e.,
$\eta \ra \infty$ even though the variation in the action
(\ref{dEH:on-shell}) is finite:
\begin{\eq}
\label{dEH:on-shell:FG}
\de I_{g(on-shell)} =\frac{1}{2}
\int_{\pa{\cal M}} d^2 x \sqrt{|\ga ^{(0)}|} T^{(2)ij}_{reg} \de
\ga^{(0)}_{ij} +O(\ep)
\end{\eq}
with
\begin{\eq}
T^{ij}_{reg} = \ep^2 T^{(2) ij}_{reg} +O(\ep^3).
\end{\eq}
So, the correct definition of the energy-momentum tensor would be
the conformally redefined energy-momentum tensor $T^{(2)ij}_{reg}$
\begin{\eq}
T^{(2) ij}_{reg} =-\f{1}{8 \pi G l} \left( \ga^{(2)ij}- \ga^{(0)ij}
Tr \ga^{(2)}
 \right) +O(\ep),
\end{\eq}
which is finite for a finite boundary metric $\ga^{(0)}_{ij}$.
 Here, it is understood that the boundary indices $(i, j, \cdots)$
in $\ga^{(n) ij}$ and $\ga^{(n)}_{ij}$ are lowered and raised by
$\ga^{(0)}_{ij}$ and its inverse $\ga^{(0) ij}$. But, note that
$T^{(2) ij}_{reg}$ depends on $\ga^{(2)ij}$, i.e., the bulk
spacetime, as well as the boundary metric $\ga^{(0) ij}$.

On the other hand, from the expansion of the Riemann tensor
\begin{\eq}
\label{Riemann:FG} {}^{(3)} {R_{i \eta}}^{k \eta}=\f{1}{l^2} \left[
{\ga^{(0)}_i}^k+ \ep^2 {(-\ga^{(2)} +3 \ga^{(4)})_i}^k \right]
+O(\ep^3)
\end{\eq}
one also finds that the first term in (\ref{tij}) vanishes, $t^{ij}
\sim -2 \be_{KL} \ep {\de_k}^i \ep^{jk \eta}/\sqrt{|\ga^{(0)}|} =0$,
up to the term of the order of $O(\ep^3)$. Then, from the second
term in the bracket `$[~~]$' in the boundary terms of (\ref{dGCS}),
one finds the boundary energy-momentum tensor
\begin{\eq}
t^{(2)ij} =\be_{KL} \f{2}{\sqrt{|\ga^{(0)}|}}\f{1}{l^2} \left[
\ep^{kj} {\ga^{(2)i}}_k +(i \lra j)  \right] +O(\ep)
\end{\eq}
with
\begin{\eq}
t^{ij}= \ep^2 t^{(2) ij} +O(\ep^3),
\end{\eq}
up to some ``non-differentiable'' boundary terms
\begin{\eq}
\label{dGCS:on-shell:non-integ}
 \de I_{GCS(on-shell)} =\frac{1}{2}
\int_{\pa{\cal M}} d^2 x \sqrt{|\ga^{(0)}|} t^{(2)ij} \de
\ga^{(0)}_{ij}-\be_{KL} \int_{\pa{\cal M}} d^2 x \Ga^k_{li} \de
\Ga^l_{kj} \ep^{ij} + O(\ep).
\end{\eq}
The unwanted boundary terms can be shown to be vanishing in the
explicit computations for the cases that I am interested in this
paper, i.e., ${\Ga^k_{li}}^{(0)}=0$ for the expansion of
$\Ga^k_{li}={\Ga^k_{li}}^{(0)}+\ep {\Ga^k_{li}}^{(2)} +\cdots$, and
so I hereafter do not consider these terms further \ci{Krau:05a}.

Then, by collecting the results, one finds the total energy-momentum
tensor for the boundary metric $\ga^{(0)}_{ij}$ as follows:
\begin{\eq}
\label{tau:total}
\tau^{(2)ij}&=&T^{(2) ij}_{reg} +t^{(2)ij} \no \\
&=&-\f{1}{8 \pi G l} \left( \ga^{(2)ij}-\ga^{(0)ij} Tr \ga^{(2)}
\right) +\f{2 \be_{KL} }{\sqrt{|\ga^{(0)}|}}\f{1}{l^2} \left[
\ep^{kj} {\ga^{(2)i}}_k +(i \lra j)  \right].
\end{\eq}
Here, I would like to note several important points in the above
result. First, there is no divergence in the action variation
(\ref{dEH:on-shell:FG}) for the Einstein-Hilbert part,
implying that the counter term (\ref{I:ct}) or its contribution in
(\ref{Tij:reg}) correctly cancels the divergent terms in the
un-regularized one. Second,
there is no additional divergences from the higher derivative term
of the gravitational Chern-Simons due to the cancelation of $\de
\ga_{ij} \ep^{ij}=0$, though this is not clear in a naive
manipulation in (\ref{dGCS}) (see Appendix {\bf B} for the details);
in other words, I do not need to consider additional counter
terms\footnote{A systematic study of the appropriate boundary terms
for some {\it arbitrary} boundary metric would be still a
challenging problem.} for this special type of higher derivative
term. Third, there is no contribution from the first term of
(\ref{tij}) in the boundary energy-momentum tensor $\tau^{ij}$ due
to some exact cancelation in the order of $O(\ep)$ when considering
${}^{(3)} {R_{i \eta}}^{k\eta} = -\ga^{kj}~ {}^{(3)} R_{i \eta j
\eta}$ of (\ref{Riemann:FG}), in contrast to a naive expectation
from ${}^{(3)} R_{i \eta k \eta} =-{l^{-2}} \ep^{-1} \ga^{(0)}_{ik}
+O(1)$. Finally, the Chern-Simons part $t^{(2)ij}$ of
(\ref{tau:total}) is {\it not} the analytic continuation from the
AdS result: Under the analytic continuation $l \ra il$, accompanied
by an additional continuation of the coordinate $\eta \ra i \eta$,
one has the usual continuation with the {\it real}-valued
energy-momentum tensor $T^{ij}_{(reg)}$ \ci{Bala:01,Sken:02}, by
considering $T^{ij}_{(reg)} \ra -i T^{ij}_{(reg)}$ and $\tau^{ij}
\ra -i \tau^{ij}$, whereas $t^{(2)ij}$ becomes {\it imaginary} in
this procedure, in contrast to the result (\ref{tau:total}). These
are resulted from the difference in the transformations, $I_{g(ct)}
\ra i I_{g(ct)}, I_{GCS} \ra I_{GCS}$\footnote{These definitions of
the actions are different from that of Ref. \ci{Witt:88}, but
consistent with the usual conventions in the holographic
renormalization \ci{Bala:01,Sken:02}} under the continuation; the
total action $I=I_g +I_{ct} +I_{GCS}$ becomes {\it complex},
$\hat{I}=I_g +I_{ct} -i I_{GCS}$ under the continuation $I \ra i
\hat{I}$, and so the Chern-Simons contributions go beyond the
analytic continuation (see Appendix {\bf C} for more details).
     \\

\begin{section}
{KdS$_3$ space and holographic energy-momentum tensor}
\end{section}

In the absence of the gravitational Chern-Simons term, a general
two-parameter family of the vacuum solution, satisfying
(\ref{eom:bulk}) with a positive cosmological constant, is known as
KdS$_3$ (three-dimensional Kerr-de Sitter) solution \ci{Park:9806}.
It would be a non-trivial task to find the general solutions for the
modified equations (\ref{eom:GCS}) with the third-order derivatives.
However, there is a trivial solution, e.g., the KdS$_3$ solution,
since it satisfies the equation (\ref{eom:bulk}) trivially from
$\nabla_{\be} {}^{(3)}{R_{\ga \rho}}^{\m \be}=0$, due to  $
{}^{(3)}R_{\m \n \al \be }=({}^{(3)} R/6 ) (g_{\m \al} g_{\n \be}
-g_{\m \n} g_{\al \be} )$ and ${}^{(3)} R=+6/l^2$.

This looks like a too-trivial situation which does not seem to have
any higher derivative effect of the gravitational Chern-Simons term.
But, actually this is not the case, as one can see, since there are
some non-trivial ``global'' effects via some shifts in the conserved
charges of the solution. So, let me introduce, first, the KdS$_3$
solution, which is given by the metric \ci{Park:9806}
\begin{eqnarray}
\label{KdS3}
 ds^2=-N^2 dt^2 +N^{-2} dr^2 +r^2 (d \phi +N^{\phi}
dt)^2
\end{eqnarray}
with
\begin{eqnarray}
\label{KdS3:N}
 N^2=8Gm-\left( \f{r}{l} \right)^2 +\f{(8 G j)^2}{4
r^2}, ~~N^{\phi}=+\f{8Gj}{2 r^2}.
\end{eqnarray}
Here, two constants of integration $m$ and $j$, which parameterize
the KdS$_3$ solution, are identified as the mass and angular
momentum of the spacetime, respectively
\ci{Klem:01,Spra:01,Bala:01}. Note the sign convention of $j$
differs from that of Refs. \ci{Park:9806,Klem:01,Spra:01,Bala:01}
but agrees with that of Ref. \ci{Klem:02}; the reason of this choice
will be clear below.

The KdS$_3$ solution (\ref{KdS3}) has one cosmological event horizon
at
\begin{eqnarray}
\label{r+}
 r_{+} =\frac{l}{\sqrt{2}} \sqrt{8Gm+\sqrt{(8 Gm)^2
+\frac{(8Gj)^2}{l^2}}}
\end{eqnarray}
and there is no black-hole event horizon. Here, there is no
additional constraint on $m$ and $j$ in order that the horizon
exists, unless $j$ vanishes: Even the negative values of $m$ are
allowed. So, in the case of $j \neq 0$, the whole mass spectrum is
continuous, ranging form $-\infty$ to $\infty$, and there is no mass
gap. For the case of $j=0$, there is no horizon when $m < 0$ and one
is left with a region which is filled with negative masses.
Moreover, if I introduce a real parameter
\begin{\eq}
r_{(-)} \equiv \frac{l}{\sqrt{2}}\sqrt{-8Gm+\sqrt{(8Gm)^2
+\frac{(8Gj)^2}{l^2}}},
\end{\eq}
 the mass and angular momentum can be
conveniently written as
\begin{eqnarray}
\label{KdS3:mj}
 m=\frac{r_+^2 -r_{(-)}^2}{8Gl^2}, ~ j =\frac{2 r_+
r_{(-)}}{8Gl}.
\end{eqnarray}
The dS$_3$ space can be identified as the case of $m=1/8G,~j=0$ in
the general KdS$_3$ solution.

The two regions separated by the cosmological horizon $r_+$ are
casually disconnected and so the cosmological event horizon acts
like as a black-hole horizon. Then, from the Gibbons-Hawking
analysis \ci{Gibb:77} this cosmological horizon produces an
isotropic background of thermal radiation with a temperature and
chemical potential
\begin{eqnarray}
\label{T:def}
 T_C=\left. \f{\hbar \kappa}{2 \pi}\right|_{r_+} =\frac{\hbar( r_{+}^2 +r_{(-)}^2)}{ 2 \pi
l^2 r_+ },~~\Omega_C=-\left.N^{\phi} \right|_{r_+}=-\f{r_{(-)}}{l
r_+ }
\end{eqnarray}
with the (positive) surface gravity function $\kappa=|\pa N^2/(2 \pa
r)|$ \footnote{In dS space, there is a subtlety in defining the
temperature, which is associated with the definition of the mass
within the cosmological horizon \ci{Klem:04}. But, here I take the
usual convention with the {\it positive} surface gravity and
temperature \ci{Gibb:77}.}.

Now, by considering the first law of thermodynamics for an arbitrary
variation `$\de$' as \footnote{ In an integrated form, one can
obtain the Smarr-type formula \ci{Smar:73}: $m=T_C S_C/2 +\Omega_C
j$.}
\begin{\eq}
\de m=T_C \de S_C + \Omega_C \de j
\end{\eq}
with $T_C$ and $\Omega_C$ as the characteristic temperature and
angular velocity of the system\footnote{If I used the sign
convention $N^{\phi}=-{8Gj}/{2 r^2}$ in (\ref{KdS3:N}), I would
obtain a wrong sign for the $\Omega_C \de j$ term. This is the
reason why I have chosen the sign convention of (\ref{KdS3:N}),
instead of Refs. \ci{Park:9806,Klem:01,Spra:01,Bala:01}. }, one can
determine the entropy of the cosmological horizon as
\begin{\eq}
S_C=\f{2 \pi r_+}{4 G \hbar},
\end{\eq}
which is the same form as the Bekenstein-Hawking entropy for black
holes. Here, the mass $m$ can be regarded as the (positive) mass
within the cosmological horizon, i.e., $r \geq r_+$, which differs
from the Gibbons-Hawking's definition
\ci{Gibb:77,Stro:0106,Klem:01,Spra:01,Klem:02} but agrees with that
of Refs. \ci{Bala:01,Myun:01,Cai:01,Cvet:01}.

On the other hand, in order to study the holographic definition of
mass, angular momentum, and its associated entropy, it is convenient
to consider the metric in the following proper radial coordinates
\ci{Bana:98b,Park:9806,Solo:05a}
\begin{eqnarray}
\label{KdS3:proper} ds^2 &=&-d \eta^2 +\f{1}{l^2}\left({r_+^2}
\mbox{sinh}^2 (\eta/l) +{r_{(-)}^2} \mbox{cosh}^2
(\eta/l) \right)dt^2  \no\\
&&~~~~~~~+ \f{1}{l^2} \left({r_+^2} \mbox{cosh}^2 (\eta/l)
+{r_{(-)}^2} \mbox{sinh}^2 (\eta/l) \right)d \phi^2 +\f{2 r_+
r_{(-)}}{l} dt d \phi
\no \\
&=&-d \eta^2 + \ep^{-1} \f{(r_+^2+r_{(-)}^2)}{4 l^2} (dt^2 + l^2 d
\phi^2) +  \f{(-r_+^2+r_{(-)}^2)}{2 l^2} (dt^2 - l^2 d \phi^2)
\no \\
&& ~~~~~~~+\f{2 r_+ r_{(-)}}{l} dt d \phi +\ep \f{r_+^2+r_{(-)}^2}{4
l^2} (dt^2 + l^2 d \phi^2).
\end{eqnarray}
Then, by comparing with (\ref{slice}), one can easily determine the
non-vanishing coefficients in the Fefferman-Graham expansion of
(\ref{FG}) as follows:
\begin{eqnarray}
\label{KdS3:FG} \ga^{(0)}_{tt}&=&l^{-2} \ga^{(0)}_{\phi \phi} =
\f{(r_+^2+r_{(-)}^2)}{4 l^2}, \no \\
\ga^{(2)}_{tt}&=&-l^{-2} \ga^{(2)}_{\phi \phi} =-
\f{(r_+^2-r_{(-)}^2)}{2 l^2}=-4Gm, \no \\
\ga^{(2)}_{t \phi}&=&
 \f{ r_+ r_{(-)}}{l}=4Gj ,  \no \\
 \ga^{(4)}_{tt}&=&l^{-2} \ga^{(4)}_{\phi \phi}
=\f{(r_+^2+r_{(-)}^2)}{4 l^2}.
\end{eqnarray}
The associated boundary energy-momentum tensor (\ref{tau:total}) can
be computed as
\begin{\eq}
\label{Tij:FG} {T^{(2)}_{tt}}_{reg}&=&-\f{1}{8 \pi G l}
\ga^{(2)}_{tt}=\f{m}{2 \pi l}, \no \\
 {T^{(2)}_{\phi
\phi}}_{reg}&=&-\f{1}{8 \pi G l} \ga^{(2)}_{\phi \phi}=-\f{ml}{2 \pi
}, \no \\
{T^{(2)}_{t \phi}}_{  reg}&=&-\f{1}{8 \pi G l} \ga^{(2)}_{t
\phi}=-\f{j}{2 \pi l}
\end{\eq}
and
\begin{\eq}
\label{tij:FG(2)} t^{(2)}_{tt}&=&\f{4 \be_{KL}}{l^2
\sqrt{|\ga^{(0)}|} } \ga^{(2)}_{t \phi} \ga^{(0)\phi \phi} \ep_{\phi
t}
=\f{16 G \be_{KL} j}{l^3}, \no \\
t^{(2)}_{\phi \phi}&=&\f{4 \be_{KL}}{l^2 \sqrt{|\ga^{(0)}|} }
\ga^{(2)}_{\phi t} \ga^{(0) tt} \ep_{t \phi } =-\f{16 G
\be_{KL} j}{l}, \no \\
\no \\
t^{(2)}_{t \phi}&=&\f{2 \be_{KL}}{l^2 \sqrt{|\ga^{(0)}|} }
\left(\ga^{(2)}_{tt} \ga^{(0) tt} \ep_{t \phi}+ \ga^{(2)}_{\phi
\phi} \ga^{(0) \phi \phi} \ep_{\phi t} \right) =\f{16 G \be_{KL}
m}{l},
\end{\eq}
where I have used $Tr (\ga^{(2)})=0$ for the solution
(\ref{KdS3:FG}) in (\ref{Tij:FG}) and $\ep_{\phi t}=\ga^{(0)}_{\phi
\phi} \ga^{(0)}_{tt} \ep^{\phi t} =+\mbox{det} \ga^{(0)}~ (\ep^{\phi
t} \equiv 1)$ with $t_{(2)ij}=\ga^{(0)}_{ik} \ga^{(0)}_{jl}
t^{(2)kl}$ in (\ref{tij:FG(2)}).

Now, according to the usual definition of mass and angular momentum
\ci{Bala:99,Krau:05a}, one obtains
\begin{\eq}
M&=&l \oint d \phi ~\tau^{(2)}_{tt} =m +\f{32 \pi G \be_{KL}
j}{l^2},
\no \\
J&=&-l \oint d \phi ~\tau^{(2)}_{t \phi} =j -32 \pi G \be_{KL} m
\end{\eq}
with $\tau^{(2)}_{ij}=T^{(2)}_{ij~reg}+t^{(2)}_{ij}$. In the absence
of the gravitational Chern-Simons contributions, these agree with
Refs. \ci{Park:9806,Bala:01}. Note that the relative signs in the
corrections terms are different. This sign difference may be
understood {\it conveniently} from the AdS case also, by considering
$(i) l \ra il,~(ii) (m, M) \ra (-m, -M),~(iii) (j,J) \ra(-j,-J)$
\ci{Park:9806}\footnote{The last step `(iii)', which is absent in
the convention of Ref. \ci{Park:9806}, is due to our definition of
(\ref{KdS3:N}) and (\ref{KdS3:mj}).} but this can not be obtained by
the conventional analytic continuation $l \ra il, \eta \ra i \eta$
as I have remarked in the previous section. Then, with these
modified mass and angular momentum, one can easily determine their
associated entropy of the cosmological horizon as
\begin{\eq}
\label{S:total}
 S=S_C +\f{16 \pi^2 \be_{KL}}{l \hbar } r_{(-)}
\end{\eq}
which satisfies the first law of thermodynamics
\begin{\eq}
\de M=T_C \de S + \Omega_C \de J
\end{\eq}
with the {\it same} $T_C$ and $\Omega_C$ as the characteristic
temperature and angular velocity of the system. The entropy
correction from the gravitational Chern-Simons term does not satisfy
the usual area law \ci{Beke:73}, similar to the BTZ black hole in
the AdS space
\ci{Solo:05a,Saho:06,Tach:06,Park:0602,Park:0608,Park:0609}. But
here, there is $no$ special meaning of the dependence of $r_{(-)}$,
in contrast to the inner horizon $r_{-}$ in the BTZ black hole.
Moreover, the dS$_3$ vacuum solution with $m=1/(8G)$ and $j=0$ has a
permanent rotation as well, in the new context,
\begin{\eq}
M=\f{1}{8 G},~ J=-4 \pi \be_{KL}.
\end{\eq}
This corresponds to the {\it chirality} in the four dimensional
cosmology with a Chern-Simons term.

Before finishing this section, I remark that the maximum mass
conjecture \ci{Bala:01,Cai:01,Ghez:01,Ghez:05} may be violated by
the modified mass $M$ since
this can be larger than that of the $dS_3$ vacuum, for
$\beta_{KL}>0$, even though the original system satisfy the
conjecture \ci{Bala:01}, i.e., $m<1/(8G)$. Moreover, the ${\bf
N}$-bound in the entropy \ci{Bous:99,Bous:00} may be violated from
the correction term in (\ref{S:total}) which does not depend on the
(outer horizon) area and can be also arbitrarily larger than that of
the $dS_3$ vacuum, for $\beta_{KL}>0$, even though it is satisfied
originally. This situation is quite similar to that of the
asymptotically dS solution with NUT charge, where the entropy is no
longer proportional to the area \ci{Clar:03,Clar:04,Mann:04}.
\\
\begin{section}
{Statistical entropy }
\end{section}

In order to compute the statistical entropy we need to know about
the holographic anomalies, first. To this end, I start by noting
that our three-dimensional gravity system would have the Weyl (or
trace) anomaly in the two-dimensional boundary, which is dictated by
the non-vanishing trace of the boundary energy-momentum tensor. And
also, I note that there is the gravitational anomaly, due to the
gravitational Chern-Simons term, which is dictated by the
non-conservation of the energy-momentum tensor or the \diff
non-invariance of the action.

As for the Weyl anomaly \ci{Ahar:99}, one usually consider the
asymptotic isometry \diff which produces some anomalous
transformations \ci{Bala:99,Bala:01}. But, an equivalent and easier
way is to consider the relation
\begin{\eq}
\label{trace:off-shell}
{\tau^{(2) i}}_i &=&{T^{(2)i}}_i +{t^{(2)i}}_i \no \\
&=&+\f{1}{8 \pi Gl} Tr (\ga^{(2)}) \no \\
&=&+\f{l R^{(0)}}{16 \pi G},
\end{\eq}
where, in the third line, I have used ${t^{(2) i}}_i \sim
\ga^{(2)}_{ik} \ep^{ik} =0$ \footnote{The non-differentiable
boundary term of (\ref{dGCS:on-shell:non-integ}) might have some
corrections in this general context, without considering the
explicit metric (\ref{KdS3:proper}). However, even in this case,
those are absent \ci{Krau:05a}.} and the bulk Einstein equation
$l^2R^{(0)}=+2~ Tr(\ga^{(2)})$ at the order of $O(\ep)$. [The
two-dimensional Ricci tensor is expanded as $R_{ij}=R^{(0)}_{ij}
+\ep R^{(2)}_{ij} + \cdots$ and $R^{(0)}={R^{(0)i}}_i$.] Then, by
comparing with the Weyl anomaly of the two-dimensional
gravity\footnote{I have introduced an {\it imaginary} factor ``$i$''
so that it agrees with the result of Ref. \ci{Park:9806}. I have
also introduced $\hbar$ in order to recover the correct $1/\hbar$
factor in the entropy \ci{Park:0608}.}
\begin{\eq}
\label{trace:central}
 {\tau^{(2) i}}_i \equiv -i \hbar \f{c -\bar{c}}{48 \pi }
R^{(0)},
\end{\eq}
one may identify the central charges
\begin{\eq}
\label{c+c:Wely_anomaly} c -\bar{c} =i \f{3 l}{G \hbar}
\end{\eq}
for the two central charges $c$ and $\bar{c}$ for the holomorphic
and anti-holomorphic sectors, generally. For the pure imaginary
central charges and $\bar{c}=c^* $, this agrees with the result of
Ref. \ci{Park:9806}. But, even for the complex valued $c$ and
$\bar{c}$, their imaginary parts are not perturbed by the
gravitational Chern-Simons term. Of course, $R^{(0)}$ vanishes
identically, i.e., ${\tau^{(2) i}}_i =0$, if one use the metric
$\ga^{(0)}_{ij}$ in the explicit solution (\ref{KdS3:proper}) so
that the computation of the central charge can not be justified for
this metric.\footnote{I thank S.Y. Nam and C. Park for drawing my
attention on this point. } But, an important point is that the
relation (\ref{trace:off-shell}) is {\it generally} valid for the
asymptotic isometries which produces non-vanishing $R^{(0)}$, and so
the computation of the central charge from (\ref{trace:central}) is
quite robust procedure.

On the other hand, the gravitational anomaly may be conventionally
computed from the variation
\begin{\eq}
\de_{\xi}I_{GCS}&=&\be_{KL} \int_{\pa{\cal M}} Tr(v d \Ga) \no \\
&\equiv& \hbar \f{c + \bar{c}}{96 \pi} \int_{\pa{\cal M}} Tr(v d
\Ga)
\end{\eq}
under a \diff $\de_{\xi}x^{\m}=-\xi^{\m}(x), ~\de_{\xi}g_{\m \n}
=v_{\m \n} +v_{\n \m}$ with ${v_{\al}}^{\be}=\pa \xi^{\al}/\pa
\xi^{\be}$,
giving
\begin{\eq}
\label{c-c:grav_anomly}
 c + \bar{c} =96 \pi \be_{KL}/\hbar.
\end{\eq}
Note that this is a genuine effect of the gravitational Chern-Simons
term, from $\de_{\xi}I_g=0$. Now, by combining
(\ref{c+c:Wely_anomaly}) and (\ref{c-c:grav_anomly}), one can get
\begin{\eq}
c &=&i \f{3 l}{2 G \hbar} + \f{48 \pi \be_{KL}}{\hbar}, \no \\
\bar{c} &=&-i \f{3 l}{2 G \hbar} + \f{48 \pi \be_{KL}}{\hbar}.
\end{\eq}
Note also that these central charges are genuine data of the
spacetimes, independently on the local structures, i.e., regardless
of the existence of the horizons. In other words, the existence of
the central charges does not necessarily mean some non-vanishing
entropy. In order to have an entropy, the system needs to have some
``energies'' and these are represented by the Virasoro generators.
So, I introduce, following Ref. \ci{Park:9806}, the zero-mode
Virasoro generators
as
\begin{\eq}
\label{L0}
L_0 &=& \f{1}{2 \hbar} (i l M +J) +\f{c}{24}, \no \\
\bar{L}_0 &=& \f{1}{2 \hbar} (-i l M +J) +\f{\bar{c}}{24}
\end{\eq}
since these satisfy the {\it usual} hermiticity condition
$L^{\dagger}_0=+\bar{L}_{(0)}$ and $c^*=\bar{c}$.
With a unitary representation of the Virasoro algebras of $L_m,
\bar{L}_m$ in the standard form, which are defined on the plane, one
can use the Cardy formula for the entropy of a CFT
\ci{Park:9806,Bala:02,Card:86,Carl:99,Park:01,Kang:04,Park:0402}
\begin{\eq}
\label{Cardy}
 S_{stat} = 2 \pi \sqrt{\f{1}{6} \left(c -24
L_{0(\min)} \right) \left(L_0-\f{c}{24} \right)} +2 \pi
\sqrt{\f{1}{6} \left(\bar{c} -24 \bar{L}_{0(\min)} \right)
\left(\bar{L}_0-\f{\bar{c}}{24} \right)},
\end{\eq}
which is real-valued and positive semi-definite, by construction. Of
course, a priori justification of this formula for the
complex-valued $c, \bar{c}$ and $L_0, \bar{L}_0$ which make the
corresponding CFT {\it non-unitary}, is still missing, but I will
just assume this formula and see what happen in our non-trivial
circumstance. Then, one can easily get ($\ga \equiv 1+i 32 \pi G
\be_{KL}/l$)
\begin{\eq}
S_{stat}&=&\f{\pi}{4 G \hbar} \left[\sqrt{-\ga^2 (r_+ -i
r_{(-)})^2}+\sqrt{-\ga^{* 2} (r_+ +i r_{(-)})^2} \right] \no \\
&=&\f{2 \pi}{4 G \hbar} \left| r_+ +\f{32 \pi G \be_{KL}}{l} r_{(-)}
\right|,
\end{\eq}
where I have chosen $L_{0(\min)}=\bar{L}_{0(\min)}=0$, as usual
\ci{Stro:97,Birm:98,Park:0403}, which corresponds to the dS$_3$
vacuum solution with $m=1/8G$ and $j=0$ in the usual context, but
with $M=1/8G, J=-4 \pi \be_{KL}$ in the new context. For a positive
coupling $\be_{KL}$ or negative $\be_{KL}$ satisfying $\be_{KL} \geq
-l r_+/(32 \pi G r_{(-)})$, this agrees exactly with the
thermodynamic entropy formula (\ref{S:total}). The disagreement for
negative couplings $\be_{KL} < -l r_+/(32 \pi G r_{(-)})$ would not
be so strange since the thermodynamic entropy (\ref{S:total})
becomes negative, whereas the statistical entropy is positive
semi-definite, by the construction.

Finally, I note that this agreement is quite non-trivial. Actually,
with the real-valued central charges $c_L =({3 l}/{2 G \hbar}) +
({48 \pi \be_{KL}}/{\hbar}),~c_R =({3 l}/{2 G \hbar}) - ({48 \pi
\be_{KL}}/{\hbar})$ as in the literatures
\ci{Stro:0106,Spra:01,Bala:01}, one can not construct a simple
formula, like (\ref{Cardy}), anymore due to the Chern-Simons
contributions. From this non-trivial agreement, the assumed formula
(\ref{Cardy}), which does not generally apply to non-unitary
theories, might have some deep meaning. However, its physical
interpretation would be quite different from that
of AdS$_{3}$, and remains to be fully understood. \\

\section*{Acknowledgments}

I would like to thank Robert B. Mann, Si-Young Nam, Chanyong Park,
and Chaiho Rim for useful correspondences. This work was supported
by the Korea Research Foundation Grant funded by Korea
Government(MOEHRD) (KRF-2007-359-C00011).

\appendix

\begin{section}
{The Gauss-Godazzi equations and the Fefferman-Graham expansion}
\end{section}

In this appendix, I consider the Gauss-Godazzi equations for the
slicings as in (\ref{slice}) and their expansions {\it \'a la}
Fefferman-Graham. I consider arbitrary dimensions with a positive or
negative cosmological constant, for the sake of generality.

I first start by considering the following slicing of the spacetime
with $(d-1)$-dimensional hypersurfaces labeled by $\eta$ ($d$
denotes the  total spacetime dimensions)
\begin{\eq}
ds^2=\mp d \eta^2 +\ga_{ij} dx^i dx^j,
\end{\eq}
where the upper (lower) sign denotes the dS (AdS) space. The
extrinsic curvature of a fixed-$\eta$ surface is defined as
$K_{ij}=+\f{1}{2} \pa_{\eta} \ga_{ij}$ and the non-vanishing
Christoffel symbols are given by
\begin{\eq}
\Ga^{\eta}_{ij} &=&\pm K_{ij},~\Ga^i_{\eta j} ={K^i}_j=\f{1}{2}
\ga^{ik} \dot{\ga}_{kj}, \no \\
\Ga^k_{ij}&=&\f{1}{2} \ga^{kl} (\pa_i \ga_{jl} +\pa_j \ga_{il}-\pa_l
\ga_{ij}),
\end{\eq}
where $\dot{\ga}_{ik} \equiv \pa_{\eta} \ga_{ik}$.

The components of the curvature tensors, known as the Gauss-Godazzi
equations,\footnote{I follow the conventions summarized in the
footnote 4. The $(d-1)$-dimensional curvatures are denoted by
${R_{ijk}}^l, \cdots,$ etc. without the superscript of `$(d-1)$' and
$D_i$ denotes the covariant derivatives with respect to the metric
$\ga_{ij}$.} are given by
\begin{\eq}
{}^{(d)} {R_{ijk}}^{\eta}&=&\pm (-D_i K_{jk} +D_j K_{ik})\no \\
&=& \pm
\f{1}{2} (-D_i \dot{\ga}_{jk} +D_j \dot{\ga}_{ik}),\no \\
{}^{(d)} {R_{i\eta k}}^{\eta}&=&\pm (\pa_{\eta} K_{ik}-{K_k}^l
K_{li} ) \no \\
&=& \pm \f{1}{2} \left(\ddot{\ga}_{ik}
-\f{1}{2} \dot{\ga}_{il} \ga^{lj} \dot{\ga}_{jk} \right),\no \\
{}^{(d)} {R_{ijk}}^{l}&=&{R_{ijk}}^{l} \mp (K_{jk} {K^l}_i -K_{ik} {K^l}_j)\no \\
&=&{R_{ijk}}^{l} \mp \f{1}{4} (\dot{\ga}_{ik} \ga^{ml}
\dot{\ga}_{jm}-\dot{\ga}_{jk} \ga^{ml} \dot{\ga}_{im}), \no \\
{}^{(d)} R_{ij}&=&R_{ij} \mp (K_{ij} ~Tr K +\pa_{\eta} K_{ij} -2 (Tr K^2)_{ij}) \no \\
&=&R_{ij} \mp \f{1}{2} \ddot{\ga}_{ij} \pm \f{1}{4} \dot{\ga}_{ij} Tr (\ga ^{-1} \dot{\ga})
\mp \f{1}{2} (\dot{\ga} \ga^{-1} \dot{\ga})_{ij}, \no \\
{}^{(d)} {R^i}_j&=&\ga^{ik}~{}^{(d)} R_{kj} = {R^i}_j\pm {K^i}_j ~Tr
K
\pm \pa_{\eta} {K^i}_{j} \no \\
&=&{R^i}_j \mp \f{1}{2} {(\ga^{-1} \ddot{\ga})^i}_j \pm \f{1}{4}
{(\ga^{-1} \dot{\ga})^i}_j Tr (\ga ^{-1} \dot{\ga})
\mp \f{1}{2} {(\ga^{-1} \dot{\ga} \ga^{-1} \dot{\ga})^i}_j, \no \\
{}^{(d)} R_{ \eta i}&=& -D_i ~Tr K +D_j {K_i}^j \no \\
&=& -\f{1}{2} D_i ~Tr (\ga^{-1} \dot{\ga}) +\f{1}{2} D_j {(\ga^{-1} \dot{\ga})^j}_i , \no \\
{}^{(d)} R_{ \eta \eta}&=& -\pa_{\eta} Tr K - Tr (K^2) \no \\
 &=&-\f{1}{2} Tr (\ga^{-1} \ddot{\ga}) +\f{1}{4} Tr (\ga^{-1}
\dot{\ga} \ga^{-1} \dot{\ga}), \no \\
{}^{(d)} R&=&{}^{(d)} {R_i}^i +{}^{(d)} {R_{\eta}}^{\eta}=R \pm
(TrK)^2 \pm Tr(K^2) \pm 2 \pa_{\eta} (Tr K) \no \\
&=&R \pm Tr(\ga^{-1} \ddot{\ga}) \pm \f{1}{4} [Tr (\ga^{-1}
\dot{\ga})]^2 \mp \f{3}{4} Tr (\ga^{-1} \dot{\ga} \ga^{-1} \dot{\ga}
).
\end{\eq}

Now, for the Fefferman-Graham expansion (\ref{FG}), the expansions
of the extrinsic curvatures are given by (\ref{K:FG}), regardless of
the sign of the cosmological constant and the spacetime dimensions.
The expansion of the Christoffel symbols ${\Ga_{ij}}^k$ is given by
\begin{\eq}
{\Ga_{ij}}^k ={\Ga_{ij}}^{(0) k} +\ep {\Ga_{ij}}^{(2) k } + \cdots,
\end{\eq}
where ${\Ga_{ij}}^{(0)k}$ is formed by $\ga^{(0)}$.

The Riemann tensors are expanded as
\begin{\eq}
{}^{(d)} R_{i \eta k\eta}&=&-\f{1}{l^2} \left[ \ep^{-1}
\ga^{(0)}_{ik} +\ga^{(2)}_{ik} + \ep (-{\ga^{(2)}}^2 +4
\ga^{(4)})_{ik}\right] +O(\ep^2)
, \no \\
{}^{(d)} R_{ijk \eta}&=&-\f{1}{l} \left[ \ep^{-1} D_i \ga^{(0)}_{jk}
- \ep D_i \ga^{(4)}_{jk} \right]-(i \lra j).
\end{\eq}
And also, by raising the indices $i,j, \cdots$ by the metric
$\ga^{ij}$,
these become
\begin{\eq}
{}^{(d)} {{R_{i \eta}}^k}_{\eta}&=&-\f{1}{l^2} \left[\ga^{(2)k}_{i}
+ \ep^2 {(-{\ga^{(2)}}^2 +3 \ga^{(4)})_i}^k  \right]+O(\ep^3), \no \\
{}^{(d)} {R^i}_{jk \eta}&=&-\f{1}{l} \left[ \ga^{(0) il} (D_l
\ga^{(0)}_{jk} -(l \lra j)) - \ep \ga^{(2) il} (D_l \ga^{(0)}_{jk}
-(l \lra j) ) \right].
\end{\eq}
Here, I note that there is an exact cancelation in the order
$O(\ep)$ of ${}^{(d)} {{R_{i \eta}}^k}_{\eta}$ by the contraction
process. This fact is crucial when considering the finite
energy-momentum tensor $t^{(2) ij}$ in (\ref{tau:total}).

The Ricci tensor and scalar are given by
\begin{\eq}
{}^{(d)} R_{\eta \eta}&=&-\f{1}{l^2} (d-1) + \f{1}{l^2} \ep^2 ~ Tr
\left({\ga^{(0)}}^2 -\ga^{(4)}\right) +O(\ep^3), \no \\
{}^{(d)} R_{\eta i}&=&\f{1}{l} \ep \left(\pa_i ~ Tr \ga^{(2)} -D_j
{\ga^{(2) j}}_i \right)-  \f{1}{l} \ep^2 \left[ \pa_i ~  Tr
\left({\ga^{(0)}}^2 -\ga^{(4)}\right) -D_j {\left({\ga^{(0)}}^2 -\ga^{(4)}\right)^j}_i
\right] +O(\ep^3), \no \\
{}^{(d)} {R^i}_{j}&=&\pm \f{1}{l^2} {\de^i}_j (d-1) + \f{1}{l^2} \ep
\left[l^2 {R^{(0)i}}_j \pm (3-d) {\ga^{(2)i}}_j \mp {\de^i}_j~ Tr
\ga^{(2)} \right] \no \\
&&-\f{1}{l^2} \ep^2 \left[-l^2 \left({R^{(2)i}}_j -\ga^{(2)ik}
R^{(0)}_{kj}\right) \pm (5-d) {\left({\ga^{(2)}}^2
-\ga^{(4)}\right)^i}_j \mp {\de^i}_j
~Tr \left({\ga^{(2)}}^2 -\ga^{(4)}\right) \right. \no \\
&&\left.\mp {\ga^{(2)i}}_j~Tr \ga^{(2)} \right]+O(\ep^3), \no \\
{}^{(d)} R&=&\pm \f{1}{l^2}d(d-1) + \f{1}{l^2} \ep \left[l^2 R^{(0)}
\pm 2(2-d)~  Tr
\ga^{(2)} \right] \no \\
&&-\f{1}{l^2} \ep^2 \left[-l^2 \left(R^{(2)} -Tr( \ga^{(2)}) \right)
\pm (-2d +6 \mp 1) ~ Tr \left({\ga^{(2)}}^2 -\ga^{(4)}\right)
\right] +O(\ep^3).
\end{\eq}
Then, by considering the bulk Einstein equation ${}^{(d)} R=\left(
\f{2d}{d-2} \right)\La$, with the cosmological constant $\La=\pm
(d-1)(d-2)/2 l^2$, one obtains the iterative (Einstein) equations
\begin{\eq}
\label{eom:0th}
R^{(0)} &=&\mp 2 (2-d) l^{-2} ~Tr \ga^{(2)},\\
R^{(2)} &=&Tr( \ga^{(2)})  \pm (-2d +6 \mp 1)l^{-2} ~ Tr
\left({\ga^{(2)}}^2 -\ga^{(4)}\right), \\
&& \vdots  \no \\
&\mbox{etc.}& \no
\end{\eq}
The first equation (\ref{eom:0th}) has been used to get the trace
anomaly of (\ref{trace:off-shell}).

\begin{section}
{Computing the energy-momentum tensor $t^{ij}$ for the gravitational
Chern-Simons term}
\end{section}

In this appendix, I compute the gravitational Chern-Simons
contributions to the energy-momentum tensor, from the term
$\int_{\pa {\cal M}} Tr ( \Ga \wedge \de \Ga )$ in (\ref{dGCS}). To
this end, I first start by noting
\begin{\eq}
\int_{\pa {\cal M}}  Tr  \left( \Ga \wedge \de \Ga \right) &=&
\int_{\pa {\cal M}} \Ga^{\al}_{\be i} \de \Ga^{\be}_{\al j}~ dx^i
\wedge
dx^j \no \\
&=&\int_{\pa {\cal M}} \left[\pm 2 {K^{l}}_i \de K_{lj} \pm
{K^{m}}_i {K^l}_j \de \ga_{ml} +\Ga_{li}^m \de \Ga_{mj}^l \right]
\ep^{ij} d^2 x ,
\end{\eq}
where the second term is canceled by $\ep^{ij}$ factor and the third
term would not contribute, as I have noted in the Sec. IV. Now, by
using the Fefferman-Graham expansion of (\ref{K:FG}) one has
\begin{\eq}
{K^{l}}_{i} \de K_{lj}&=&\f{1}{l^2}\ep^{-1} \de
\ga^{(0)}_{ij}-\f{1}{l^2} {\ga^{(2)l}}_i \de \ga^{(0)}_{lj}  \no
\\
&&+\f{1}{l^2} \ep \left[ -\de \ga^{(4)}_{ij}+{\left(
(\ga^{(2)})^2-\ga^{(4)}\right)^{l}}_{i} \de \ga^{(0)}_{lj} \right]
+O (\ep^2),
\end{\eq}
which has a divergence in the first term. But, by considering the
factor $\ep^{ij}$ again, this term and the first term in the bracket
do not contribute. So, I finally have
\begin{\eq}
\de I_{GCS(on-shell)} &=&\mp 2 \be_{KL} \int_{\pa {\cal M}} d^2 x ~
\ep^{ij}~ {K^{l}}_i
\de K_{lj} \no \\
&=&\mp 2 \be_{KL} \int_{\pa {\cal M}} d^2 x  \left[ -\f{1}{l^2}
 \ep^{lj} {\ga^{(2)i}}_l \de \ga^{(0)}_{ij} \right. \no \\
 &&\left.+\f{1}{l^2}   \ep~\ep^{lj}
{\left( (\ga^{(2)})^2-\ga^{(4)}\right)^{i}}_{l} \de \ga^{(0)}_{ij}
\right] +O(\ep^2),
\end{\eq}
which produces the boundary energy-momentum tensor of
(\ref{dGCS:on-shell:non-integ}):
\begin{\eq}
\label{tij:general}
 t^{(2) ij} =\pm \be_{KL} \f{2
}{\sqrt{|\ga^{(0)}|}}\f{1}{l^2} \left[ \ep^{kj} {\ga^{(2)i}}_k +(i
\lra j)  \right] +O(\ep)
\end{\eq}
with $t^{ij}=\ep^2 t^{(2)ij} +O(\ep^3)$. Here, I note that the
overall sign differs from that of (5.6) in Ref. \ci{Krau:05a} and
this results a different sign in the gravitational Chern-Simons
corrections to the mass and angular
momentum. (See the next section for the details.)\\

\begin{section}
{Comparative computation of the conserved charges in dS/AdS and the
analytic continuation}
\end{section}

In this appendix, I consider the computation of the conserved
charges by comparing the dS and AdS cases, for completeness, and the
analytic continuation between them. This will clarify a sign
difference with the literatures.

To this end, I first start by considering the Einstein-Hilbert
action, with the Gibbons-Hawking's boundary term, in arbitrary
$d$-dimensions and with arbitrary cosmological constants $\La=\pm
(d-1)(d-2)/2 l^2$, is
\begin{eqnarray}
\label{EH:dS/AdS} I_{g}&=&\frac{1}{16 \pi G} \int_{\cal M} d^3 x
\sqrt{|g|} \left( {}^{(d)}R -2 \La \right)\mp \frac{1}{8 \pi G}
\int_{\pa {\cal M}}
d^{d-1} x \sqrt{|\ga|} Tr K \no \\
 &=&\frac{1}{16 \pi G} \int_{\cal M}
d^{d-1} x d\eta   \sqrt{|\ga|} \left(R \pm(Tr K)^2 \pm Tr (K^2)-2
\La \right).
\end{eqnarray}
(My conventions are the same as those of Refs.
\ci{Krau:05a,Noji:99}, but different from those of Refs.
\ci{Solo:05a,Stro:0106,Spra:01,Bala:01,Bala:99,Sken:02}.) Then, the
variation produces, when applying the bulk equations of motion,
\begin{\eq}
\label{dEH:on-shell:d-dim} \de I_{g(on-shell)} =\frac{1}{2}
\int_{\pa{\cal M}} d^{d-1} x \sqrt{|\ga|} T^{ij} \de \ga_{ij}
\end{\eq}
with
\begin{\eq}
T^{ij} =\pm \f{1}{8 \pi G} \left(K^{ij} -Tr(K) \ga^{ij}  \right).
\end{\eq}
Now, with the counter terms \ci{Krau:99,Bala:99,Bala:01,Noji:01b},
which can be fixed by the locality and general covariance,
\begin{\eq}
\label{I:ct:d-dim}
 I_{ct}=\pm \frac{1}{8 \pi G } \int_{\pa{\cal M}} d^{d-1} x
\sqrt{|\ga|} \left[ \f{d-2}{l} \mp \f{l}{2 (d-3)} R \right]
\end{\eq}
one obtains  the regulated energy-momentum tensor
\begin{\eq}
\label{Tij:reg:d-dim}
 T^{ij}_{reg}=T^{ij}\pm \f{1}{8 \pi G }\left[ \f{d-2}{l} \ga^{ij} \pm \f{l}{d-3}\left(R^{ij}
 -\f{1}{2} R \ga^{ij} \right) \right].
\end{\eq}
In the Fefferman-Graham expansion, this  becomes
\begin{\eq}
\label{Tij:reg:d-dim:FG}
 T^{ij}_{reg}=\mp \f{1}{8 \pi G } \f{\ep^2}{l} \left[ (4-d)
 \ga^{(2)ij}-\ga^{(0)ij} Tr \ga^{(2)}
   \right] + O(\ep^3).
\end{\eq}
For the three-dimensional case, i.e., $d=3$, this reduces to
\begin{\eq}
\label{Tij:reg:3-dim:FG}
 T^{ij}_{reg}=\mp \f{1}{8 \pi G } \f{\ep^2}{l} \left[
 \ga^{(2)ij}-\ga^{(0)ij} Tr \ga^{(2)}
  \right] + O(\ep^3),
\end{\eq}
which agrees with (\ref{Tij:reg:FG}). Here, I note that these
results can be also obtained by the analytic continuation $l \ra il,
R_{ij} \ra R_{ij}, I_g \ra i I_g, I_{ct} \ra i I_{ct}$, and
$T^{ij}_{(reg)} \ra - i T^{ij}_{(reg)}$ \ci{Bala:01,Sken:02}.

By combining $t^{(2)ij}$ of (\ref{tij:general}) from the
gravitational Chern-Simons term, the total boundary energy-momentum
tensor in $d=3$ is given by
\begin{\eq}
\tau^{ij}= \ep^2 \tau^{(2) ij} +O(\ep^3)
\end{\eq}
with
\begin{\eq}
\label{tau(2):(a)dS}
 \tau^{(2)ij} =\mp \f{1}{8 \pi G l} \left[
\ga^{(2)ij}- \ga^{(0)ij} Tr \ga^{(2)}  \right] \mp \f{2 \be_{KL}
}{l^2 \sqrt{|\ga^{(0)}|}} \left[ \ep^{jk} {\ga^{(2)i}}_k +(i \lra j)
\right].
\end{\eq}
Here, note that the Chern-Simons contributions can {\it not} be
analytically continued, in contrast to $T^{ij}_{reg}$ parts since
the Chern-Simons contributions become {\it imaginary} under the
continuation $\tau^{ij} \ra -i \tau^{ij}$ with $l \ra il$.

I also note the appearance of $1/\sqrt{|\ga^{(0)}|}$ factor in the
second term, in contrast to the literatures \ci{Krau:05a,Solo:05a}.
Then, from the Fefferman-Graham expansion of the metric
\ci{Bana:98b,Park:9806,Solo:05a}
\begin{eqnarray}
\label{KdS3/BTZ:FG} \ga^{(0)}_{tt}&=&\pm l^{-2} \ga^{(0)}_{\phi
\phi} =\pm
\f{(r_+^2 \pm r_{(-)}^2)}{4 l^2}, \no \\
\ga^{(2)}_{tt}&=&\mp l^{-2} \ga^{(2)}_{\phi \phi} =\mp
\f{(r_+^2 \mp r_{(-)}^2)}{2 l^2}= \mp 4Gm, \no \\
\ga^{(2)}_{t \phi}&=&
 \pm \f{ r_+ r_{(-)}}{l}=\pm 4Gj ,  \no \\
 \ga^{(4)}_{tt}&=&\pm l^{-2} \ga^{(4)}_{\phi \phi}
=\pm  \f{(r_+^2 \pm r_{(-)}^2)}{4 l^2},
\end{eqnarray}
one can find the total boundary energy-momentum tensors
\begin{\eq}
\label{tij:total:(a)dS}
\tau^{(2)}_{tt} &=&\f{m}{2
\pi l} \pm \f{16 G \be_{KL} j}{l^3}, \no \\
\tau^{(2)}_{\phi \phi}&=&\mp \f{ml}{2 \pi } -\f{16 G
\be_{KL} j}{l}, \no \\
\tau^{(2)}_{t \phi}&=&-\f{j}{2 \pi l }+\f{16 G \be_{KL} m}{l},
\end{\eq}
where I have used the usual definition $\ep_{\phi t}=+\mbox{det}
\ga^{(0)}~ (\ep^{\phi t} \equiv 1)$ and $\mbox{det} \ga^{(0)}
/|\mbox{det} \ga^{(0)}|=\mbox{sign}( \ga^{(0)} )=\pm 1$ with
$\mbox{det} \ga^{(0)}=\pm (r_+^2 \pm r_{(-)}^2 )^2/(4 l)^2$.

Then, the conserved charges becomes
\begin{\eq}
M&=&l \oint d \phi ~\tau^{(2)}_{tt} =m \pm \f{32 \pi G \be_{KL}
j}{l^2},
\no \\
J&=&-l \oint d \phi ~\tau^{(2)}_{t \phi} =j -32 \pi G \be_{KL} m .
\end{\eq}
These final results agree with those of Refs.
\ci{Bala:99,Bala:01,Krau:05a,Solo:05a} and (1) with $\hb=-32 \pi G
\be_{KL}/l$ \ci{Park:0608}, but the details in the computations are
different. The asymptotic metric $\ga^{(0)}_{ij}$ has the arbitrary
conformal factors ${(r_+^2 \pm r_{(-)}^2)}/{4 l^2}$, which are
functions of $m$ and $j$, in contrast to the literatures. There is
no difference in the computation of $T^{(2)}_{ij ~reg}$ since there
is no contribution of $\ga^{(0)}_{ij}$ for the solution
(\ref{KdS3/BTZ:FG}). However, the conformal factors are crucial in
the computation of $t^{(2)}_{ij}$ in order to get the correct
results of (\ref{tij:total:(a)dS}): In Refs. \ci{Krau:05a,Solo:05a}
(similarly also in \ci{Bala:99,Bala:01}), $\ga^{(0)}_{ij}$ has been
considered as $\ga^{(0)}_{tt}=- l^{-2} \ga^{(0)}_{  \phi \phi} =-1$
with $\mbox{det} \ga^{(0)} =-l^2$, even with the {\it same} higher
metric $\ga^{(n)}$ as in (\ref{KdS3/BTZ:FG})\footnote{This implies
that the generic KdS$_3$/BTZ metric can not be ``smoothly''
transformed into the forms in Refs.
\ci{Krau:05a,Solo:05a,Bala:99,Bala:01}. In the context of the
substraction procedure \ci{Gibb:77b,Brow:93}, this corresponds to
choose the reference spacetime as $\ga^{(0)}_{ij}$ in
(\ref{KdS3/BTZ:FG}), which depends on the mass and angular
momentum.}, and some strange normalization $\ep^{ \phi t }= l$, in
contrast to the usual one as above. On the contrary, with the
correct factor of $1/\sqrt{|\ga^{(0)}|}$ in the gravitational
Chern-Simons part in (\ref{tau(2):(a)dS}) or (\ref{tij:FG(2)}), one
can find the correct results, even for the asymptotic metric
$\ga^{(0)}$ with arbitrary masses and angular momenta.\\

\newcommand{\J}[4]{#1 {\bf #2} #3 (#4)}
\newcommand{\andJ}[3]{{\bf #1} (#2) #3}
\newcommand{\AP}{Ann. Phys. (N.Y.)}
\newcommand{\MPL}{Mod. Phys. Lett.}
\newcommand{\NP}{Nucl. Phys.}
\newcommand{\PL}{Phys. Lett. }
\newcommand{\PR}{Phys. Rev. }
\newcommand{\PRL}{Phys. Rev. Lett.}
\newcommand{\PTP}{Prog. Theor. Phys.}
\newcommand{\CQG}{Class. Quant, Grav.}
\newcommand{\hep}[1]{ hep-th/{#1}}
\newcommand{\hepg}[1]{ gr-qc/{#1}}
\newcommand{\bi}{ \bibitem}

\end{document}